\documentstyle[aps,prb,epsf,multicol,amsmath]{revtex}

\begin{document}
\draft

\newcommand{\cd}{\hat c} 
\newcommand{\cdd}{\Hat{\Hat{c}}}

\title{Construction of size-consistent effective Hamiltonians for systems with arbitrary Hilbert space}
\author{Arnd H\"{u}bsch, Matthias Vojta\cite{Vojta}, and Klaus W. Becker}
\address{
  Institut f\"{u}r Theoretische Physik,
  Technische Universit\"{a}t Dresden, D-01062 Dresden, Germany 
}
\maketitle

\begin{abstract}
Effective Hamiltonians are usually constructed by using canonical transformations or projection techniques. In contrast to this, we present a method for systems with arbitrary Hilbert space based on the introduction of cumulants. Cumulants guarantee size consistency, a property that is not always evident in other treatments. As a nontrivial example of use the derived method is applied to the strong-coupling limit of the half-filled Hubbard model on a general lattice in arbitrary spatial dimension for which the fourth-order expansion in $t/U$ of the effective Hamiltonian is derived.
\end{abstract}

\widetext
\begin{multicols}{2}
\narrowtext


\section{Introduction}

The construction of effective Hamiltonians has proved very useful in investigating complicated physical systems, since in this way the number of degrees of freedom can be reduced. Conventional methods for such a construction are based on canonical transformations or projection technique. Often the construction is carried out in an approximate way, usually by second-order perturbation theory. The transformation of the Hubbard model \cite{Hubbard,Hubbard2,Hubbard3} into the $t$-$J$ model \cite{Chao1,Gros,Zhang} is a well-known example.

The method of canonical transformations was introduced by Schrieffer and Wolff \cite{Schrieffer} for transferring the Anderson Hamiltonian \cite{Anderson2} into the Kondo model \cite{Kondo}. This approach can be sketched as follows. For a given Hamiltonian of the form 
\begin{eqnarray}
  {\cal H}&=&{\cal H}_{0}+\epsilon{\cal H}_{1}\label{G1}
\end{eqnarray}
one searches for a generator $S$ of the transformation
\begin{eqnarray}
  {\cal H}'&=&e^{-\epsilon S}{\cal H}e^{\epsilon S}\label{G2}
\end{eqnarray}
under the condition that the new Hamiltonian ${\cal H}'$ does not contain linear terms in $\epsilon$. It can be shown that ${\cal H}_{1}=\left[ S,{\cal H}_{0}\right]$ is the desired conditional equation for the generator $S$. If this is solved for $S$ and inserted into \eqref{G2}, one obtains the second-order result of ${\cal H}'$. Since this construction involves commutation operations with the original Hamiltonian ${\cal H}$, the effective Hamiltonian ${\cal H}'$ scales with the size of the system. The size consistency is important, since otherwise the results for extensive quantities like ground-state energy or magnetization would prove inconsistent. For a size-consistent calculation of higher-order terms, this simple method must be extended. This can be done either by a step-by-step transformation \cite{MacDonald} or by the method of continuous unitary transformations \cite{Wegner}. However, these approaches are very formal and therefore not transparent.

In contrast, the conventional projection method \cite{Fulde} is based on a division of the Hilbert space into two subspaces: the ${\rm U}_{P}$ subspace in which one is interested and the complementary ${\rm U}_{Q}$ subspace to be projected out with the respective projection operators  $P$ and $Q={\bf 1}-P$. Using the projection operators $P$ and $Q$ the Schr\"{o}dinger equation is split into two parts, and the states of the ${\rm U}_{Q}$ subspace can be eliminated. Thus one obtains a Schr\"{o}dinger-type equation $\left({\cal H}_{P}^{{\rm eff}}-E\right)\left|\psi_{P}\right\rangle=0$ for the ${\rm U}_{P}$ subspace where $\left|\psi_{P}\right\rangle$ denotes a state of the ${\rm U}_{P}$ subspace. The effective Hamiltonian is given by
\begin{eqnarray}
  {\cal H}_{P}^{{\rm eff}}&=&P{\cal H}P-P{\cal H}Q\frac{1}{Q{\cal H}Q-E}Q{\cal H}P.\label{G5}
\end{eqnarray}
If the two subspaces interact only weakly, the states of the subspaces may be separated by a typical energy $\Delta$. In this case one selects the projection operators $P$ and $Q$ for a Hamiltonian of the form \eqref{G1} in such a way that the unperturbed Hamiltonian ${\cal H}_{0}$ cannot provide transitions between the subspaces. By replacing $Q{\cal H}Q-E$ by $\Delta$, one obtains the second order in $\epsilon$ of the effective Hamiltonian ${\cal H}_{P}^{{\rm eff}}$. For a size-consistent computation of higher orders it is not sufficient to expand Eq.~\eqref{G5} for small $Q{\cal H}_{1}Q$. In addition the energy $E$ and the state $|\psi_{P}\rangle$ must be expanded consistently. The resulting equations for the separate $\epsilon$-orders contain eigenvalue corrections of different orders. Therefore these equations must be solved successively \cite{Elbaz}, and it is difficult to obtain closed expressions for higher orders. 

In contrast to the usual projection method the cumulant approach \cite{Becker1,Becker2,Becker3} preserves size consistency of extensive variables and is therefore a suitable tool for the construction of effective Hamiltonians. A previously developed cumulant approach \cite{Polatsek1} can only be applied to systems consisting of two interacting subsystems. This method is generalized here for arbitrary systems. 

The paper is organized as follows. In Sec.~\ref{formalism} the general cumulant approach is presented. As a test for the derived method in Sec.~\ref{Hubbard} the effective Hamiltonian up to the fourth order in $t/U$ for the half-filled Hubbard model is calculated. The conclusions are presented in Sec.~\ref{Conclusions}. Finally, a detailed discussion of generalized cumulants is given in the Appendix.


\section{Cumulant approach}\label{formalism}

The cumulant approach \cite{Becker1,Becker2,Becker3} has established itself as a powerful technique of many-body theory, which makes the investigation of static and dynamical ground-state properties of weakly and strongly correlated systems possible. It is known from classical statistical mechanics\cite{Ursell,Mayer} that size consistency is attained by expressing extensive quantities in terms of cumulants, i.e. a cumulant expression for an extensive variable scales with the size of the system independent of further approximations. In the standard diagram technique size consistency is ensured in any approximation by considering linked diagrams only.\cite{Negele} But a diagrammatic approach is usually based on Wick's theorem\cite{Wick} which is only applicable, if the dominant part of the Hamiltonian is a single-particle operator. Therefore, the diagrammatic description is restricted to weakly correlated systems in which the electron-electron interaction may be treated pertubatively.

In the following we propose a cumulant approach for the construction of effective Hamiltonians for arbitrary systems. The method presented here is based on a perturbational approach for the Hamiltonian ${\cal H}={\cal H}_{0}+{\cal H}_{1}$. The Hilbert space of the unperturbed Hamiltonian ${\cal H}_{0}$ is split into the low-energy part ${\rm U}_{P}$ and the high-energy part ${\rm U}_Q$ with the respective projection operators $P$ and $Q={\bf 1}-P$. The states of these two subspaces are assumed to be separated by an energy difference. Now we want to construct an effective Hamiltonian for the ${\rm U}_P$ subspace. For that purpose the expression for the effective Hamiltonian from Ref.~\ref{Polatsek1}
\begin{eqnarray}
  {\cal H}^{\rm eff}_{a}&=&-\frac{1}{\beta}\ln\left[\frac{1}{Z_{b}}{\rm tr}_{b}\left( e^{-\beta {\cal H}}\right)\right]\label{G7}
\end{eqnarray}
must be discussed first. Here $\beta$ is the inverse temperature. In this case a system with two subsets of degrees of freedom $a$ and $b$ was considered, i.e. the Hilbert space of the system is a product space of $a$ and $b$. In Eq.~\eqref{G7} $Z_{b}$ and ${\rm tr}_{b}$ denote the partition function and the trace of the subspace $b$. Eq.~\eqref{G7} has been derived from the partition function of the whole system by separation of the trace. Therefore the recently introduced cumulant method for the construction of effective Hamiltonians \cite{Polatsek1} is limited to systems on product spaces. 

Now Eq.~\eqref{G7} is generalized by replacing the trace by a projection onto the relevant ${\rm U}_P$ subspace
\begin{eqnarray}
  {\cal H}_{P}^{\rm eff}&=&-\frac{1}{\beta}P\ln\left( e^{-\beta{\cal H}}\right)_{P}P,\label{G8}
\end{eqnarray}
where $(\dots)_{P}$ denotes $P(\dots)P$. This generalization is the key step in the present construction. In Eq.~\eqref{G7} a part of the degrees of freedom is integrated out by the trace in the subspace $b$. In anlogy to this, insignificant degrees of freedom are removed in Eq.~\eqref{G8} by a $P$ projection onto the ${\rm U}_P$ subspace.

In order to transform Eq.~\eqref{G8} into a cumulant expression we introduce generalized cumulants
\begin{eqnarray}
  \left( \prod_{i=1}^{N}X_{i}^{\nu_{i}}\right)^{C}_{P}&=&\left[\prod_{i=1}^{N}\left(\frac{\partial}{\partial \xi_{i}}\right)^{\nu_{n}}\right]\ln\left(\prod_{i=1}^{N}e^{\xi_{i}X_{i}}\right)_{P}\bigg|_{\xi_{i}=0\;\forall i}.\nonumber\\
  &&\label{G8a}
\end{eqnarray}
For a detailed discussion of generalized cumulants see Appendix \ref{A1}. We define the operator function
\begin{eqnarray}
  f(\lambda)&=&P\ln\left( e^{-\lambda {\cal H}}\right)_{P}P\label{G8b}\\
  &=&\left( e^{-\lambda {\cal H}}-{\bf 1}\right)_{P}^{C},\label{G9}
\end{eqnarray}
where Eq.~\eqref{G9} follows from the definition \eqref{G8b} by using series expansions. The expression of the effective Hamiltonian can be rewritten
\begin{eqnarray}
  {\cal H}_{P}^{\rm eff}&=&-\frac{1}{\beta}f(\beta).
\end{eqnarray}
For an actual calculation of the effective Hamiltonian as formulated here, it is often more practical to start from the Laplace transform  $F(z)$ of the operator function $f(\lambda)$ 
\begin{eqnarray}
  F(z)&=&-\int\limits_{-\infty}^{0}e^{z\lambda}f(\lambda)d\lambda\label{G9a}\\
      &=&-\frac{1}{z^{2}}\left[ ({\cal H})_{P}^{C}+\left( {\cal H}\frac{1}{z-{\cal H}}{\cal H}\right)_{P}^{C}\right].\label{G10}
\end{eqnarray}
It can be shown that the second term of Eq.~\eqref{G10} does not contribute to the zeroth and the first order of ${\cal H}_{1}$ and therefore
\begin{eqnarray}
  F(z)&=&-\frac{1}{z^{2}}\left[ ({\cal H})_{P}^{C}+z\left({\cal H}_{1}\frac{1}{z-{\cal H}}{\cal H}_{1}\frac{1}{z-{\cal H}_{0}}\right)_{P}^{C}+\right.\nonumber\\
  &&\quad+\left. z\left( {\cal H}_{0}\frac{1}{z-{\cal H}}{\cal H}_{1}\frac{1}{z-{\cal H}_{0}}{\cal H}_{1}\frac{1}{z-{\cal H}_{0}}\right)_{P}^{C}\right]\label{G11}\\
  &=&-\frac{1}{z^{2}}\left({\cal H}\right)^{C}_{P}-\sum\limits_{n=2}^{\infty}\left(\left[\frac{1}{z-{\cal H}_{0}}{\cal H}_{1}\right]^{n}\frac{1}{z-{\cal H}_{0}}\right)_{P}^{C}.\nonumber\\
  &&\label{G12}
\end{eqnarray}
Now we have obtained the sought after perturbation series for the effective Hamiltonian and the separate cumulant expansions corresponding to the perturbation orders in ${\cal H}_{1}$. Note that the derived cumulant expression is exact and applicable at arbitrary temperature since no restrictive assumptions were used. It can be shown that the previously developed cumulant method \cite{Polatsek1} is a particular case of our formalism.

In the following we want to discuss a special situation of the method derived above which also provides an interesting approximation. If all states of the relevant ${\rm U}_{P}$ subspace have the same eigenvalue $E_{P}$ of the unperturbed Hamiltonian ${\cal H}_{0}$, the formalism can be simplified, because in this case all cumulant expressions of the forms  $(\dots{\cal H}_{0})_{P}^{C}$ and $({\cal H}_{0}\dots)_{P}^{C}$ vanish, and Eq.~\eqref{G10} can be rewritten
\begin{eqnarray}
  F(z)&=&-\frac{1}{z^{2}}\left[ ({\cal H})_{P}^{C}+\left({\cal H}_{1}\frac{1}{z-{\cal H}}{\cal H}_{1}\right)_{P}^{C}\right]\label{G12a}\\
  &=&-\frac{1}{z^{2}}\left[ ({\cal H})_{P}^{C}+\left({\cal H}_{1}\frac{1}{z-L_{0}-{\cal H}_{1}}{\cal H}_{1}\right)_{P}^{C}\right].\label{G13}
\end{eqnarray}
$L_{0}$ is the Liouville operator with respect to ${\cal H}_{0}$ which is defined by $L_{0}{\cal A}=\left[{\cal H}_{0},{\cal A}\right]$ for any operators ${\cal A}$. Eq. \eqref{G13} can be shown by transforming the second term of \eqref{G12a} into a series of cumulant expressions and using
\begin{eqnarray}
  \left({\cal H}_{1}{\cal H}^{n}{\cal H}_{1}\right)_{P}^{C}&=&\left({\cal H}_{1}(L_{0}+{\cal H}_{1})^{n}{\cal H}_{1}\right)_{P}^{C}.\label{G14b}
\end{eqnarray}
Note that Eq.~\eqref{G14b} is proved by using the unit operator ${\bf 1}=e^{-\xi{\cal H}_{0}}e^{\xi{\cal H}_{0}}$, the identity \cite{Becker1} $e^{\lambda{\cal H}}{\cal A} e^{-\lambda{\cal H}_{0}}=e^{\lambda(L_{0}+{\cal H}_{1})}{\cal A}$ and the definition of generalized cumulant expressions \eqref{G8a}. If we want to restrict ourself to zero temperature, the inverse Laplace transform of Eq.~\eqref{G13} is easly performed and
\begin{eqnarray}
  \lefteqn{{\cal H}_{P}^{\rm eff}(\beta\rightarrow\infty)\;=}&&\nonumber\\
  &=&({\cal H}_{0})^{C}_{P}+\lim_{z\rightarrow 0}\left({\cal H}_{1}\frac{1}{z-L_{0}-{\cal H}_{1}}{\cal H}_{1}\right)_{P}^{C}\label{G15b}
\end{eqnarray}
is then obtained for the effective Hamiltonian. In the case ${\cal H}_{1}$ is small relative to ${\cal H}_{0}$, it is straightforward to expand the effective Hamiltonian in perturbation series. From Eq.~\eqref{G15b} it follows that
\begin{eqnarray}
  \lefteqn{{\cal H}_{P}^{\rm eff}(\beta\rightarrow\infty)\;=}&&\nonumber\\
  &=&({\cal H}_{0})^{C}_{P}+\lim_{z\rightarrow 0}\left\{\sum\limits_{n=0}^{\infty}\left({\cal H}_{1}\left[\frac{1}{z-L_{0}}{\cal H}_{1}\right]^{n}\right)_{P}^{C}\right\}.\label{G15}
\end{eqnarray}
As an alternative to perturbation theory, the evaluation of the effective Hamiltonian \eqref{G13} can also be done in the framework of the Mori-Zwanzig projection method \cite{Zwanzig,Mori} as was shown in Ref.~\ref{Polatsek1}. In the following we want to restrict ourselves to the perturbation result \eqref{G15}.


\section{Application to the half-filled Hubbard Model}\label{Hubbard}

We now apply the method described above to derive an effective Hamiltonian for the strong-coupling limit of the half-filled Hubbard model up to fourth order in $t/U$. As it turned out in the past this derivation is highly nontrivial; recently Stein\cite{Stein} has used it as a test for Wegner's method of continuous transformations.\cite{Wegner} Therefore it is reasonable to utilize the cumulant approach presented here for re-examination of this expansion.

The Hamiltonian of the Hubbard model \cite{Hubbard,Hubbard2,Hubbard3} on a general lattice in arbitrary spatial dimension is given by
\begin{eqnarray}
  {\cal H}&=&{\cal H}_{0}+{\cal H}_{1},\label{G15a}\\
  {\cal H}_{0} &=& \frac{1}{2}\; U\sum_{i,\sigma}n_{i,\sigma}n_{i,-\sigma},\label{G16}\\
  {\cal H}_{1}&=&t\sum_{i,j,\sigma}D_{i,j}\;c_{i,\sigma}^{\dagger}c_{j,\sigma},\label{G17}
\end{eqnarray}
where $c_{i,\sigma}^{\dagger}$ and $c_{i,\sigma}$ are fermion creation and annihilation operators for an electron with spin $\sigma$ on site $i$. $t$ is the hopping integral, $U$ denotes the Coulomb repulsion between electrons on the same site and $n_{i,\sigma}=c_{i,\sigma}^{\dagger}c_{i,\sigma}$ is the occupation-number operator for electrons with spin $\sigma$ on site $i$. All informations about the lattice are contained in the real hopping matrix $D_{i,j}$. The Hubbard model is one of the simplest models which describes Coulomb interaction ${\cal H}_{0}$ and kinetic energy ${\cal H}_{1}$.

The expansion of the half-filled Hubbard model around the limit of strong coupling, $U\gg t$, has a long history, because this limit makes several simplifications possible. Half filling means that the electron number equals the number of lattice sites. Before the presentation of the Hubbard model, Anderson \cite{Anderson1} showed that such a system is described by an effective spin Hamiltonian, the so called Heisenberg antiferromagnet \cite{Manousakis}. This Hamiltonian is in accordance with the second order $(t/U)$ expansion of the half-filled Hubbard model.

The calculation of higher orders is often based on the perturbation theory of Kato \cite{Kato}. Thus Klein and Seitz \cite{Klein1,Seitz1,Klein2} obtained the sixth-order spin interaction for the linear chain, and Bulaevski's result \cite{Bulaevskii} of the fourth perturbation order for the half-filled Hubbard model in more than one dimension was corrected by Takahashi \cite{Takahashi1}.

Another approach for the derivation of effective Hamiltonians is based on unitary transformations. Harris and Lange \cite{Harris} used such a transformation to obtain the second-order perturbation. A transformation for calculating higher orders was introduced by Chao, Spa{\l}ek und Ole\'s \cite{Chao1,Chao2,Chao3}. Beyond second order their results are faulty, because in accordance with Eq.~\eqref{G2} the transformed Hamiltonian is constructed using a low-order approximation of the generator $S$ and so in higher orders terms appear which mix the different Hubbard bands  \cite{Oles,MacDonald2}. A correct transformation algorithm for deriving a perturbation series of the Hubbard Hamiltonian was proposed by  MacDonald, Girvin and Yoshioka \cite{MacDonald}. Thus they could calculate the fourth order of the effective Hamiltonian. Stein \cite{Stein} used Wegner's method \cite{Wegner} of continuous unitary transformations to obtain the fourth-order perturbation of the effective Hamiltonian.

We now apply the method described in Sec.~\ref{formalism} to construct an effective Hamiltonian for the half-filled Hubbard model \cite{Hubbard,Hubbard2,Hubbard3}. At half filling all states without doubly occupied lattice sites have the lowest eigenvalue of the unperturbed Hamiltonian ${\cal H}_{0}$. These states form the low-energy subspace with associated projection operator $P$. This subspace is degenerate with respect to ${\cal H}_{0}$. Terms with odd powers in the perturbation ${\cal H}_{1}$ vanish due to the particle-hole symmetry. This is in accordance with a general theorem shown by Takahashi \cite{Takahashi1}. Therefore from Eq.~\eqref{G15} follows
\begin{eqnarray}
   {\cal H}_{\rm eff}&=&\lim_{z\rightarrow 0}\left\{\sum\limits_{n=0}^{\infty}\left({\cal H}_{1}\left[\frac{1}{z-L_{0}}{\cal H}_{1}\right]^{2n+1}\right)_{P}^{C}\right\}\label{G18}
\end{eqnarray}
for the effective Hamiltonian, where the terms in the sum correspond to the perturbation orders. The first term of Eq.~\eqref{G15} vanishes because the eigenvalue of the unperturbed Hamiltonian ${\cal H}_{0}$ of the states without double occupied lattice sites is $0$.

For the calculation of the cumulant expressions it is profitable to decompose ${\cal H}_{1}$
\begin{eqnarray}
  {\cal H}_{1}&=&\sum_{k=1}^{4}h_{k},\qquad L_{0}h_{k}\;=\;\Delta_{k}h_{k}\label{G21}
\end{eqnarray}
into eigenoperators $h_{k}$ of the Liouville operator $L_{0}$ with eigenvalues $\Delta_{k}$. The searched decomposition
\begin{eqnarray}
  h_{1}&=&t\sum_{i,j,\sigma}D_{i,j}\cd^{\dagger}_{i,\sigma}\cd_{j,\sigma}\qquad\Delta_{1}\; =\; 0 \label{G22}\\
  h_{2}&=&t\sum_{i,j,\sigma}D_{i,j}\cd^{\dagger}_{i,\sigma}\cdd_{j,\sigma}\qquad\Delta_{2}\; =\; -U \label{G23}\\
  h_{3}&=&t\sum_{i,j,\sigma}D_{i,j}\cdd^{\dagger}_{i,\sigma}\cd_{j,\sigma}\qquad\Delta_{3}\; =\; U \label{G2_24}\\
  h_{4}&=&t\sum_{i,j,\sigma}D_{i,j}\cdd^{\dagger}_{i,\sigma}\cdd_{j,\sigma}\qquad\Delta_{4}\; =\; 0.\label{G25}
\end{eqnarray}
can be derived by introduction of Hubbard operators
\begin{eqnarray}
  \cd^{\dagger}_{i,\sigma}&=&c^{\dagger}_{i,\sigma}(1-n_{i,-\sigma})\label{G19}\\
  \cdd^{\dagger}_{i,\sigma}&=&c^{\dagger}_{i,\sigma}n_{i,-\sigma}\;.\label{G20}
\end{eqnarray}
Note that $\cd^{\dagger}$ describes transitions from empty to singly occupied sites, whereas $\cdd^{\dagger}$ describes transitions from singly to doubly occupied sites.

The second perturbation order is given by the first term in the sum of Eq.~\eqref{G18}. Using the identity
\begin{eqnarray}
  \lefteqn{\left(X\frac{1}{z-L_{0}}h_{1}\dots h_{N}\right)_{P}^{C}\;=}&&\nonumber\\
  &=&\frac{1}{z-(\Delta_{1}+\dots+\Delta_{N})}\left(X h_{1}\dots h_{N}\right)_{P}^{C}\label{G26}
\end{eqnarray}
which can be proved by series expansions \cite{Polatsek1} we obtain
\begin{eqnarray}
  {\cal H}_{\rm eff}\bigg|_{\rm 2nd\,ord.}&=&\lim_{z\to 0}\left[ \sum_{k=1}^{4}\frac{1}{z-\Delta_{k}}\left( {\cal H}_{1}h_{k}\right)_{P}^{C}\right].\label{G27}
\end{eqnarray}
One calculates the cumulant expression in accordance with the definition \eqref{G8a} by using series expansions (for more details see Appendix \ref{A1} and Eq.~\eqref{G45}). Due to the properties of the $h_{k}$, expressions of the form $(h_{k})_{P}$ vanish. Only the term with $k=3$ contributes so that the sum can be evaluated 
\begin{eqnarray}
  {\cal H}_{\rm eff}\bigg|_{\rm 2nd\,ord.}&=&-\frac{1}{U}\left( {\cal
  H}_{1}{\cal H}_{1}\right)_{P}.\label{G28}
\end{eqnarray}
Since ${\cal H}_{1}$ describes a hopping process, Eq.~\eqref{G28} has a clear interpretation. Due to the Pauli principle, contributions only result from different spins on the sites involved in the process. Time reversed initial states can be treated commonly by a spin summation. Therefore only two processes contribute to the second-order perturbation and one can write symbolically
\begin{eqnarray}
  {\cal H}_{\rm eff}\bigg|_{\rm 2nd\,ord.}&=&-\frac{1}{U}\left[\mbox{\setlength{\unitlength}{0.4ex}
\begin{picture}(10,12)
  \multiput(0,-6)(10,0){2}{\line(0,1){15}}
  \multiput(0,-6)(0,7.5){3}{\line(1,0){10}}
  \put(0,-4.5){\makebox(5,5)[t]{$\uparrow$}}
  \put(5,-4.5){\makebox(5,5)[t]{$\downarrow$}}
  \put(0,3){\makebox(5,5)[t]{$\uparrow$}}
  \put(5,3){\makebox(5,5)[t]{$\downarrow$}}
\end{picture}}+\mbox{\setlength{\unitlength}{0.4ex}
\begin{picture}(10,12)
  {\small 
  \multiput(0,-6)(10,0){2}{\line(0,1){15}}
  \multiput(0,-6)(0,7.5){3}{\line(1,0){10}}
  \put(0,-4.5){\makebox(5,5)[t]{$\Downarrow$}}
  \put(5,-4.5){\makebox(5,5)[t]{$\Uparrow$}}
  \put(0,3){\makebox(5,5)[t]{$\uparrow$}}
  \put(5,3){\makebox(5,5)[t]{$\downarrow$}}
  }
\end{picture}
}\right]\label{G29}
\end{eqnarray}
for Eq.~\eqref{G28}. The diagrams in Eq.~\eqref{G29} are to be interpreted as hopping processes where the initial state is located above and the final state of the process is below. The arrows with double lines $\Uparrow$ or $\Downarrow$ indicate spins which have been flipped.

Using creation and annihilation operators Eq.~\eqref{G29} can be transformed into an operator equation. Thereby the lattice site summations, the spin summation and the pre-factors of the hopping processes in accordance with Eq.~\eqref{G17} have to be considered. The sign of the diagrams results from the process execution and the fermionic anti-commutation relations. From Eq.~\eqref{G29} follows directly
\begin{eqnarray}
  {\cal H}_{\rm eff}\bigg|_{\rm 2nd\,ord.}&=&-\frac{t^{2}}{U}\sum\limits_{i,j,\sigma}D_{i,j}D_{j,i}\nonumber\\
  &&\quad\times\left[ n_{i,\sigma}n_{j,-\sigma}+c^{\dagger}_{i,\sigma}c_{i,-\sigma}c_{j,\sigma}c^{\dagger}_{j,-\sigma}\right].\label{G30}
\end{eqnarray}
Using spin operators
\begin{eqnarray}
  {\bf S}_{i}&=&\frac{1}{2}\sum\limits_{\sigma,\sigma'}c^{\dagger}_{i,\sigma}\,{\boldsymbol\sigma}_{\sigma,\sigma'}\, c_{i,\sigma'}\label{G31}
\end{eqnarray}
Eq.~\eqref{G30} can be written as follows
\begin{eqnarray}
  {\cal H}_{\rm eff}\bigg|_{\rm 2nd\,ord.}&=&2\,\frac{t^{2}}{U}\sum\limits_{i,j}D_{i,j}D_{j,i}\left[ {\bf S}_{i}\cdot{\bf S}_{j}-\frac{1}{4}\right],\label{G32}
\end{eqnarray}
where ${\boldsymbol\sigma}_{\sigma,\sigma'}$ denotes the vector of the Pauli spin matrices.

For the calculation of the fourth perturbation order in $t/U$, the second term in the sum of Eq.~\eqref{G18} has to be evaluated. The cumulant expression can be simplified again by using Eq.~\eqref{G26}
\begin{eqnarray}
  {\cal H}_{\rm eff}\bigg|_{\rm 4th\,ord.}&=&-\sum\limits_{j,k,l,m=1}^{4}\frac{1}{\Delta_{k}+\Delta_{l}+\Delta_{m}}\frac{1}{\Delta_{l}+\Delta_{m}}\frac{1}{\Delta_{m}}\nonumber\\
  &&\qquad\times\left(h_{j}h_{k}h_{l}h_{m}\right)_{P}^{C}.\label{G33}
\end{eqnarray} 
The cumulant expression has to be calculated by using Eq.~\eqref{G42}. Due to the properties of the $h_{k}$ terms of the form $(h_{k})_{P}$ vanish and therefore Eq.~\eqref{G33} can be transformed into 
\begin{eqnarray}
  {\cal H}_{\rm eff}\bigg|_{\rm 4th\,ord.}&=&\frac{1}{U^{3}}\left[\left({\cal H}_{1}{\cal H}_{1}\right)_{P}\left({\cal H}_{1}{\cal H}_{1}\right)_{P}+\right.\nonumber\\
  &&\qquad\left. -U\left({\cal H}_{1}{\cal H}_{1}Q\frac{1}{{\cal H}_{0}}Q{\cal H}_{1}{\cal H}_{1}\right)_{P}\right].\label{G34}
\end{eqnarray}
Now the fourth order should be transformed to a spin Hamiltonian too. In accordance with Eq.~\eqref{G16} all states have the eigenvalue $aU$ of the unperturbed Hamiltonian ${\cal H}_{0}$ where $a$ denotes the number of doubly occupied lattice sites of this state. Since ${\cal H}_{1}$ describes a hopping process, Eq.~\eqref{G34} can be evaluated transparently. For this purpose, it is appropriate to classify the contributing processes and to treat these classes in separate manner. 

Note that only connected diagrams contribute to the effective Hamiltonian. This fact results from the effect of cumulant expressions and secures the size-consistency. The processes to be considered can be classified according to the number of involved sites. As an example, the contribution of the processes with three involved sites to the effective Hamiltonian will be calculated in the following. In analogy to this computation, the processes with two and four involved sites can be considered too.

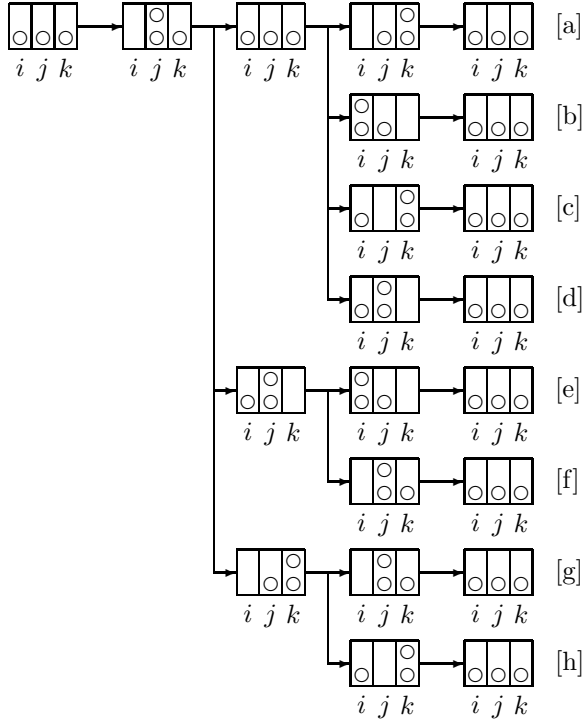
\begin{figure}
\begin{center}
\setlength{\unitlength}{0.4ex}
\begin{picture}(135,165)
  \multiput(5,150)(5,0){4}{\line(0,1){10}}
  \multiput(30,150)(5,0){4}{\line(0,1){10}}
  \multiput(55,150)(5,0){4}{\line(0,1){10}}
  \multiput(80,150)(5,0){4}{\line(0,1){10}}
  \multiput(105,150)(5,0){4}{\line(0,1){10}}
  \multiput(80,130)(5,0){4}{\line(0,1){10}}
  \multiput(105,130)(5,0){4}{\line(0,1){10}}
  \multiput(80,110)(5,0){4}{\line(0,1){10}}
  \multiput(105,110)(5,0){4}{\line(0,1){10}}
  \multiput(80,90)(5,0){4}{\line(0,1){10}}
  \multiput(105,90)(5,0){4}{\line(0,1){10}}
  \multiput(55,70)(5,0){4}{\line(0,1){10}}
  \multiput(80,70)(5,0){4}{\line(0,1){10}}
  \multiput(105,70)(5,0){4}{\line(0,1){10}}
  \multiput(80,50)(5,0){4}{\line(0,1){10}}
  \multiput(105,50)(5,0){4}{\line(0,1){10}}
  \multiput(55,30)(5,0){4}{\line(0,1){10}}
  \multiput(80,30)(5,0){4}{\line(0,1){10}}
  \multiput(105,30)(5,0){4}{\line(0,1){10}}
  \multiput(80,10)(5,0){4}{\line(0,1){10}}
  \multiput(105,10)(5,0){4}{\line(0,1){10}}
  \multiput(5,150)(25,0){5}{\line(1,0){15}}
  \multiput(5,160)(25,0){5}{\line(1,0){15}}
  \multiput(80,130)(25,0){2}{\line(1,0){15}}
  \multiput(80,140)(25,0){2}{\line(1,0){15}}
  \multiput(80,110)(25,0){2}{\line(1,0){15}}
  \multiput(80,120)(25,0){2}{\line(1,0){15}}
  \multiput(80,90)(25,0){2}{\line(1,0){15}}
  \multiput(80,100)(25,0){2}{\line(1,0){15}}
  \multiput(55,70)(25,0){3}{\line(1,0){15}}
  \multiput(55,80)(25,0){3}{\line(1,0){15}}
  \multiput(80,50)(25,0){2}{\line(1,0){15}}
  \multiput(80,60)(25,0){2}{\line(1,0){15}}
  \multiput(55,30)(25,0){3}{\line(1,0){15}}
  \multiput(55,40)(25,0){3}{\line(1,0){15}}
  \multiput(80,10)(25,0){2}{\line(1,0){15}}
  \multiput(80,20)(25,0){2}{\line(1,0){15}}
  \multiput(5,143)(25,0){5}{\makebox(5,5)[t]{$i$}}
  \multiput(80,123)(25,0){2}{\makebox(5,5)[t]{$i$}}
  \multiput(80,103)(25,0){2}{\makebox(5,5)[t]{$i$}}
  \multiput(80,83)(25,0){2}{\makebox(5,5)[t]{$i$}}
  \multiput(55,63)(25,0){3}{\makebox(5,5)[t]{$i$}}
  \multiput(80,43)(25,0){2}{\makebox(5,5)[t]{$i$}}
  \multiput(55,23)(25,0){3}{\makebox(5,5)[t]{$i$}}
  \multiput(80,3)(25,0){2}{\makebox(5,5)[t]{$i$}}
  \multiput(10,143)(25,0){5}{\makebox(5,5)[t]{$j$}}
  \multiput(85,123)(25,0){2}{\makebox(5,5)[t]{$j$}}
  \multiput(85,103)(25,0){2}{\makebox(5,5)[t]{$j$}}
  \multiput(85,83)(25,0){2}{\makebox(5,5)[t]{$j$}}
  \multiput(60,63)(25,0){3}{\makebox(5,5)[t]{$j$}}
  \multiput(85,43)(25,0){2}{\makebox(5,5)[t]{$j$}}
  \multiput(60,23)(25,0){3}{\makebox(5,5)[t]{$j$}}
  \multiput(85,3)(25,0){2}{\makebox(5,5)[t]{$j$}}
  \multiput(15,143)(25,0){5}{\makebox(5,5)[t]{$k$}}
  \multiput(90,123)(25,0){2}{\makebox(5,5)[t]{$k$}}
  \multiput(90,103)(25,0){2}{\makebox(5,5)[t]{$k$}}
  \multiput(90,83)(25,0){2}{\makebox(5,5)[t]{$k$}}
  \multiput(65,63)(25,0){3}{\makebox(5,5)[t]{$k$}}
  \multiput(90,43)(25,0){2}{\makebox(5,5)[t]{$k$}}
  \multiput(65,23)(25,0){3}{\makebox(5,5)[t]{$k$}}
  \multiput(90,3)(25,0){2}{\makebox(5,5)[t]{$k$}}
  \multiput(20,155)(25,0){4}{\vector(1,0){10}}
  \multiput(95,15)(0,20){7}{\vector(1,0){10}}
  \multiput(70,35)(0,40){2}{\vector(1,0){10}}
  \put(50,35){\line(0,1){120}}
  \put(75,95){\line(0,1){60}}
  \multiput(75,15)(0,40){2}{\line(0,1){20}}
  \multiput(75,15)(0,40){4}{\vector(1,0){5}}
  \put(75,115){\vector(1,0){5}}
  \multiput(50,35)(0,40){2}{\vector(1,0){5}}
  \multiput(107.5,12.5)(0,20){8}{\circle{3}}
  \multiput(112.5,12.5)(0,20){8}{\circle{3}}
  \multiput(117.5,12.5)(0,20){8}{\circle{3}}
  \multiput(7.5,152.5)(5,0){3}{\circle{3}}
  \multiput(57.5,152.5)(5,0){3}{\circle{3}}
  \multiput(37.5,152.5)(0,5){2}{\circle{3}}
  \put(42.5,152.5){\circle{3}}
  \multiput(92.5,152.5)(0,5){2}{\circle{3}}
  \put(87.5,152.5){\circle{3}}
  \multiput(82.5,132.5)(0,5){2}{\circle{3}}
  \put(87.5,132.5){\circle{3}}
  \put(82.5,112.5){\circle{3}}
  \multiput(92.5,112.5)(0,5){2}{\circle{3}}
  \put(82.5,92.5){\circle{3}}
  \multiput(87.5,92.5)(0,5){2}{\circle{3}}
  \multiput(82.5,72.5)(0,5){2}{\circle{3}}
  \put(87.5,72.5){\circle{3}}
  \multiput(87.5,52.5)(0,5){2}{\circle{3}}
  \put(92.5,52.5){\circle{3}}
  \multiput(87.5,32.5)(0,5){2}{\circle{3}}
  \put(92.5,32.5){\circle{3}}
  \put(82.5,12.5){\circle{3}}
  \multiput(92.5,12.5)(0,5){2}{\circle{3}}
  \put(57.5,72.5){\circle{3}}
  \multiput(62.5,72.5)(0,5){2}{\circle{3}}
  \put(62.5,32.5){\circle{3}}
  \multiput(67.5,32.5)(0,5){2}{\circle{3}}
  \put(125,150){\makebox(10,10)[l]{[a]}}
  \put(125,130){\makebox(10,10)[l]{[b]}}
  \put(125,110){\makebox(10,10)[l]{[c]}}
  \put(125,90){\makebox(10,10)[l]{[d]}}
  \put(125,70){\makebox(10,10)[l]{[e]}}
  \put(125,50){\makebox(10,10)[l]{[f]}}
  \put(125,30){\makebox(10,10)[l]{[g]}}
  \put(125,10){\makebox(10,10)[l]{[h]}}
\end{picture}
\end{center}
\caption{
Processes of the fourth-order perturbation with three involved lattice sites. The spin labelling is dropped. Note that the geometrical position of the sites $i$, $j$, $k$ is fixed by the hopping matrix $D_{i,j}$.
}
\label{ord_4_3}
\end{figure}

Fig.~\ref{ord_4_3} shows the processes with three involved sites which are possible in the case of half filling. The selected first hopping process does not influence the results due to the lattice site summations. Note that the geometrical position of the sites $i$, $j$, $k$ is fixed by the hopping matrix $D_{i,j}$. In accordance with the number of doubly occupied sites after two hopping processes, the processes can be divided into two groups. In the cases [a]-[d] respectively [e]-[h] one notices no respectively one doubly occupied sites. Therefore the processes [a]-[d] contribute to the first and [e]-[h] contribute to the second term of Eq.~\eqref{G34}. This fact influences the sign of the process diagrams.

\begin{figure}
\begin{center}
\setlength{\unitlength}{0.4ex}
\begin{picture}(125,85)
  \multiput(5,70)(5,0){4}{\line(0,1){10}}
  \multiput(30,70)(5,0){4}{\line(0,1){10}}
  \multiput(55,70)(5,0){4}{\line(0,1){10}}
  \multiput(80,70)(5,0){4}{\line(0,1){10}}
  \multiput(105,70)(5,0){4}{\line(0,1){10}}
  \multiput(5,30)(5,0){4}{\line(0,1){10}}
  \multiput(30,30)(5,0){4}{\line(0,1){10}}
  \multiput(55,30)(5,0){4}{\line(0,1){10}}
  \multiput(80,30)(5,0){4}{\line(0,1){10}}
  \multiput(105,30)(5,0){4}{\line(0,1){10}}
  \multiput(105,50)(5,0){4}{\line(0,1){10}}
  \multiput(105,10)(5,0){4}{\line(0,1){10}}
  \multiput(5,70)(25,0){5}{\line(1,0){15}}
  \multiput(5,80)(25,0){5}{\line(1,0){15}}
  \multiput(5,30)(25,0){5}{\line(1,0){15}}
  \multiput(5,40)(25,0){5}{\line(1,0){15}}
  \multiput(105,50)(0,10){2}{\line(1,0){15}}
  \multiput(105,10)(0,10){2}{\line(1,0){15}}
  \multiput(5,63)(25,0){5}{\makebox(5,5)[t]{$i$}}
  \multiput(5,23)(25,0){5}{\makebox(5,5)[t]{$i$}}
  \multiput(105,3)(0,40){2}{\makebox(5,5)[t]{$i$}}
  \multiput(10,63)(25,0){5}{\makebox(5,5)[t]{$j$}}
  \multiput(10,23)(25,0){5}{\makebox(5,5)[t]{$j$}}
  \multiput(110,3)(0,40){2}{\makebox(5,5)[t]{$j$}}
  \multiput(15,63)(25,0){5}{\makebox(5,5)[t]{$k$}}
  \multiput(15,23)(25,0){5}{\makebox(5,5)[t]{$k$}}
  \multiput(115,3)(0,40){2}{\makebox(5,5)[t]{$k$}}
  \multiput(20,75)(25,0){4}{\vector(1,0){10}}
  \multiput(20,35)(25,0){4}{\vector(1,0){10}}
  \put(100,15){\line(0,1){20}}
  \put(100,55){\line(0,1){20}}
  \multiput(100,15)(0,40){2}{\vector(1,0){5}}
  \multiput(7.5,72.5)(50,0){3}{\vector(0,1){5}}
  \multiput(36.5,72.5)(55,0){2}{\vector(0,1){5}}
  \multiput(38.5,77.5)(55,0){2}{\vector(0,-1){5}}
  \multiput(12.5,77.5)(50,0){2}{\vector(0,-1){5}}
  \multiput(87.5,77.5)(25,0){2}{\vector(0,-1){5}}
  \multiput(17.5,77.5)(25,0){3}{\vector(0,-1){5}}
  \put(117.5,77.5){\vector(0,-1){5}}
  \put(105,52.5){\makebox(5,5)[t]{$\Downarrow$}}
  \put(115,52.5){\makebox(5,5)[t]{$\Uparrow$}}
  \put(112.5,57.5){\vector(0,-1){5}}
  \put(7.5,37.5){\vector(0,-1){5}}
  \put(12.5,32.5){\vector(0,1){5}}
  \put(17.5,37.5){\vector(0,-1){5}}
  \put(36.5,32.5){\vector(0,1){5}}
  \put(38.5,37.5){\vector(0,-1){5}}
  \put(42.5,37.5){\vector(0,-1){5}}
  \put(57.5,32.5){\vector(0,1){5}}
  \put(62.5,37.5){\vector(0,-1){5}}
  \put(67.5,37.5){\vector(0,-1){5}}
  \put(87.5,37.5){\vector(0,-1){5}}
  \put(91.5,32.5){\vector(0,1){5}}
  \put(93.5,37.5){\vector(0,-1){5}}
  \put(105,32.5){\makebox(5,5)[t]{$\Uparrow$}}
  \put(110,32.5){\makebox(5,5)[t]{$\Downarrow$}}
  \put(117.5,37.5){\vector(0,-1){5}}
  \put(107.5,17.5){\vector(0,-1){5}}
  \put(110,12.5){\makebox(5,5)[t]{$\Downarrow$}}
  \put(115,12.5){\makebox(5,5)[t]{$\Uparrow$}}
\end{picture}
\end{center}
\caption{
Schematic representation of the diagrams contributing to process [a] of Fig.~\ref{ord_4_3}.
}
\label{sa_4_3}
\end{figure}
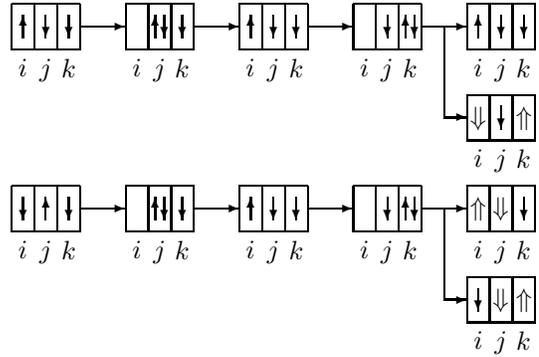

\begin{figure}
\begin{center}
\setlength{\unitlength}{0.4ex}
\begin{picture}(125,85)
  \multiput(5,70)(5,0){4}{\line(0,1){10}}
  \multiput(30,70)(5,0){4}{\line(0,1){10}}
  \multiput(55,70)(5,0){4}{\line(0,1){10}}
  \multiput(80,70)(5,0){4}{\line(0,1){10}}
  \multiput(105,70)(5,0){4}{\line(0,1){10}}
  \multiput(5,30)(5,0){4}{\line(0,1){10}}
  \multiput(30,30)(5,0){4}{\line(0,1){10}}
  \multiput(55,30)(5,0){4}{\line(0,1){10}}
  \multiput(80,30)(5,0){4}{\line(0,1){10}}
  \multiput(105,30)(5,0){4}{\line(0,1){10}}
  \multiput(105,50)(5,0){4}{\line(0,1){10}}
  \multiput(105,10)(5,0){4}{\line(0,1){10}}
  \multiput(5,70)(25,0){5}{\line(1,0){15}}
  \multiput(5,80)(25,0){5}{\line(1,0){15}}
  \multiput(5,30)(25,0){5}{\line(1,0){15}}
  \multiput(5,40)(25,0){5}{\line(1,0){15}}
  \multiput(105,50)(0,10){2}{\line(1,0){15}}
  \multiput(105,10)(0,10){2}{\line(1,0){15}}
  \multiput(5,63)(25,0){5}{\makebox(5,5)[t]{$i$}}
  \multiput(5,23)(25,0){5}{\makebox(5,5)[t]{$i$}}
  \multiput(105,3)(0,40){2}{\makebox(5,5)[t]{$i$}}
  \multiput(10,63)(25,0){5}{\makebox(5,5)[t]{$j$}}
  \multiput(10,23)(25,0){5}{\makebox(5,5)[t]{$j$}}
  \multiput(110,3)(0,40){2}{\makebox(5,5)[t]{$j$}}
  \multiput(15,63)(25,0){5}{\makebox(5,5)[t]{$k$}}
  \multiput(15,23)(25,0){5}{\makebox(5,5)[t]{$k$}}
  \multiput(115,3)(0,40){2}{\makebox(5,5)[t]{$k$}}
  \multiput(20,75)(25,0){4}{\vector(1,0){10}}
  \multiput(20,35)(25,0){4}{\vector(1,0){10}}
  \put(100,15){\line(0,1){20}}
  \put(100,55){\line(0,1){20}}
  \multiput(100,15)(0,40){2}{\vector(1,0){5}}
  \put(7.5,72.5){\vector(0,1){5}}
  \put(12.5,77.5){\vector(0,-1){5}}
  \put(17.5,77.5){\vector(0,-1){5}}
  \put(36.5,72.5){\vector(0,1){5}}
  \put(38.5,77.5){\vector(0,-1){5}}
  \put(42.5,77.5){\vector(0,-1){5}}
  \put(57.5,77.5){\vector(0,-1){5}}
  \put(61.5,72.5){\vector(0,1){5}}
  \put(63.5,77.5){\vector(0,-1){5}}
  \put(81.5,72.5){\vector(0,1){5}}
  \put(83.5,77.5){\vector(0,-1){5}}
  \put(87.5,77.5){\vector(0,-1){5}}
  \put(107.5,72.5){\vector(0,1){5}}
  \put(112.5,77.5){\vector(0,-1){5}}
  \put(117.5,77.5){\vector(0,-1){5}}
  \put(105,52.5){\makebox(5,5)[t]{$\Downarrow$}}
  \put(112.5,57.5){\vector(0,-1){5}}
  \put(115,52.5){\makebox(5,5)[t]{$\Uparrow$}}
  \put(7.5,37.5){\vector(0,-1){5}}
  \put(12.5,32.5){\vector(0,1){5}}
  \put(17.5,37.5){\vector(0,-1){5}}
  \put(36.5,32.5){\vector(0,1){5}}
  \put(38.5,37.5){\vector(0,-1){5}}
  \put(42.5,37.5){\vector(0,-1){5}}
  \put(57.5,37.5){\vector(0,-1){5}}
  \put(61.5,32.5){\vector(0,1){5}}
  \put(63.5,37.5){\vector(0,-1){5}}
  \put(81.5,32.5){\vector(0,1){5}}
  \put(83.5,37.5){\vector(0,-1){5}}
  \put(87.5,37.5){\vector(0,-1){5}}
  \put(105,32.5){\makebox(5,5)[t]{$\Uparrow$}}
  \put(110,32.5){\makebox(5,5)[t]{$\Downarrow$}}
  \put(117.5,37.5){\vector(0,-1){5}}
  \put(110,12.5){\makebox(5,5)[t]{$\Downarrow$}}
  \put(115,12.5){\makebox(5,5)[t]{$\Uparrow$}}
  \put(107.5,17.5){\vector(0,-1){5}}

\end{picture}
\end{center}
\caption{
Schematic representation of the diagrams contributing to process [e] of Fig.~\ref{ord_4_3}.
}
\label{se_4_3}
\end{figure}
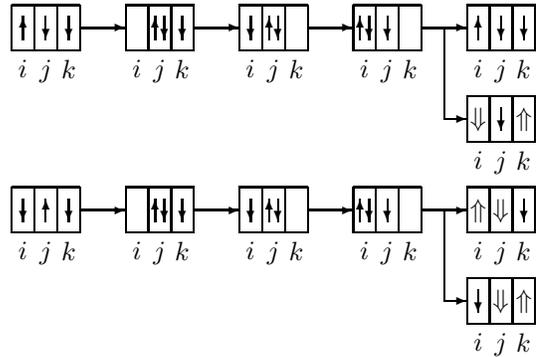

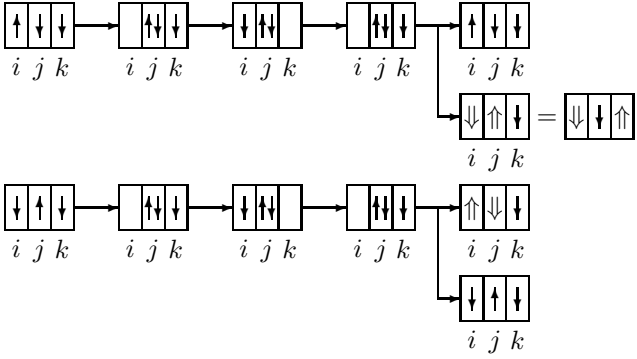
\begin{figure}
\begin{center}
\setlength{\unitlength}{0.4ex}
\begin{picture}(143,85)
  \multiput(0,70)(5,0){4}{\line(0,1){10}}
  \multiput(25,70)(5,0){4}{\line(0,1){10}}
  \multiput(50,70)(5,0){4}{\line(0,1){10}}
  \multiput(75,70)(5,0){4}{\line(0,1){10}}
  \multiput(100,70)(5,0){4}{\line(0,1){10}}
  \multiput(0,30)(5,0){4}{\line(0,1){10}}
  \multiput(25,30)(5,0){4}{\line(0,1){10}}
  \multiput(50,30)(5,0){4}{\line(0,1){10}}
  \multiput(75,30)(5,0){4}{\line(0,1){10}}
  \multiput(100,30)(5,0){4}{\line(0,1){10}}
  \multiput(100,50)(5,0){4}{\line(0,1){10}}
  \multiput(100,10)(5,0){4}{\line(0,1){10}}
  \multiput(0,70)(25,0){5}{\line(1,0){15}}
  \multiput(0,80)(25,0){5}{\line(1,0){15}}
  \multiput(0,30)(25,0){5}{\line(1,0){15}}
  \multiput(0,40)(25,0){5}{\line(1,0){15}}
  \multiput(100,50)(0,10){2}{\line(1,0){15}}
  \multiput(100,10)(0,10){2}{\line(1,0){15}}
  \multiput(0,63)(25,0){5}{\makebox(5,5)[t]{$i$}}
  \multiput(0,23)(25,0){5}{\makebox(5,5)[t]{$i$}}
  \multiput(100,3)(0,40){2}{\makebox(5,5)[t]{$i$}}
  \multiput(5,63)(25,0){5}{\makebox(5,5)[t]{$j$}}
  \multiput(5,23)(25,0){5}{\makebox(5,5)[t]{$j$}}
  \multiput(105,3)(0,40){2}{\makebox(5,5)[t]{$j$}}
  \multiput(10,63)(25,0){5}{\makebox(5,5)[t]{$k$}}
  \multiput(10,23)(25,0){5}{\makebox(5,5)[t]{$k$}}
  \multiput(110,3)(0,40){2}{\makebox(5,5)[t]{$k$}}
  \multiput(15,75)(25,0){4}{\vector(1,0){10}}
  \multiput(15,35)(25,0){4}{\vector(1,0){10}}
  \put(95,15){\line(0,1){20}}
  \put(95,55){\line(0,1){20}}
  \multiput(95,15)(0,40){2}{\vector(1,0){5}}
  \put(116.5,50.5){\makebox(5,5)[t]{=}}
  \multiput(123,50)(5,0){4}{\line(0,1){10}}
  \multiput(123,50)(0,10){2}{\line(1,0){15}}
  \put(2.5,72.5){\vector(0,1){5}}
  \put(7.5,77.5){\vector(0,-1){5}}
  \put(12.5,77.5){\vector(0,-1){5}}
  \put(31.5,72.5){\vector(0,1){5}}
  \put(33.5,77.5){\vector(0,-1){5}}
  \put(37.5,77.5){\vector(0,-1){5}}
  \put(52.5,77.5){\vector(0,-1){5}}
  \put(56.5,72.5){\vector(0,1){5}}
  \put(58.5,77.5){\vector(0,-1){5}}
  \put(81.5,72.5){\vector(0,1){5}}
  \put(83.5,77.5){\vector(0,-1){5}}
  \put(87.5,77.5){\vector(0,-1){5}}
  \put(102.5,72.5){\vector(0,1){5}}
  \put(107.5,77.5){\vector(0,-1){5}}
  \put(112.5,77.5){\vector(0,-1){5}}
  \put(100,52.5){\makebox(5,5)[t]{$\Downarrow$}}
  \put(105,52.5){\makebox(5,5)[t]{$\Uparrow$}}
  \put(112.5,57.5){\vector(0,-1){5}}
  \put(123,52.5){\makebox(5,5)[t]{$\Downarrow$}}
  \put(130.5,57.5){\vector(0,-1){5}}
  \put(133,52.5){\makebox(5,5)[t]{$\Uparrow$}}
  \put(2.5,37.5){\vector(0,-1){5}}
  \put(7.5,32.5){\vector(0,1){5}}
  \put(12.5,37.5){\vector(0,-1){5}}
  \put(31.5,32.5){\vector(0,1){5}}
  \put(33.5,37.5){\vector(0,-1){5}}
  \put(37.5,37.5){\vector(0,-1){5}}
  \put(52.5,37.5){\vector(0,-1){5}}
  \put(56.5,32.5){\vector(0,1){5}}
  \put(58.5,37.5){\vector(0,-1){5}}
  \put(81.5,32.5){\vector(0,1){5}}
  \put(83.5,37.5){\vector(0,-1){5}}
  \put(87.5,37.5){\vector(0,-1){5}}
  \put(100,32.5){\makebox(5,5)[t]{$\Uparrow$}}
  \put(105,32.5){\makebox(5,5)[t]{$\Downarrow$}}
  \put(112.5,37.5){\vector(0,-1){5}}
  \put(102.5,17.5){\vector(0,-1){5}}
  \put(107.5,12.5){\vector(0,1){5}}
  \put(112.5,17.5){\vector(0,-1){5}}
\end{picture}
\end{center}
\caption{
Schematic representation of the diagrams contributing to process [f] of Fig.~\ref{ord_4_3}.
}
\label{sf_4_3}
\end{figure}

\begin{figure}
\begin{center}
\setlength{\unitlength}{0.4ex}
\begin{picture}(143,85)
  \multiput(0,70)(5,0){4}{\line(0,1){10}}
  \multiput(25,70)(5,0){4}{\line(0,1){10}}
  \multiput(50,70)(5,0){4}{\line(0,1){10}}
  \multiput(75,70)(5,0){4}{\line(0,1){10}}
  \multiput(100,70)(5,0){4}{\line(0,1){10}}
  \multiput(0,30)(5,0){4}{\line(0,1){10}}
  \multiput(25,30)(5,0){4}{\line(0,1){10}}
  \multiput(50,30)(5,0){4}{\line(0,1){10}}
  \multiput(75,30)(5,0){4}{\line(0,1){10}}
  \multiput(100,30)(5,0){4}{\line(0,1){10}}
  \multiput(100,50)(5,0){4}{\line(0,1){10}}
  \multiput(100,10)(5,0){4}{\line(0,1){10}}
  \multiput(0,70)(25,0){5}{\line(1,0){15}}
  \multiput(0,80)(25,0){5}{\line(1,0){15}}
  \multiput(0,30)(25,0){5}{\line(1,0){15}}
  \multiput(0,40)(25,0){5}{\line(1,0){15}}
  \multiput(100,50)(0,10){2}{\line(1,0){15}}
  \multiput(100,10)(0,10){2}{\line(1,0){15}}
  \multiput(0,63)(25,0){5}{\makebox(5,5)[t]{$i$}}
  \multiput(0,23)(25,0){5}{\makebox(5,5)[t]{$i$}}
  \multiput(100,3)(0,40){2}{\makebox(5,5)[t]{$i$}}
  \multiput(5,63)(25,0){5}{\makebox(5,5)[t]{$j$}}
  \multiput(5,23)(25,0){5}{\makebox(5,5)[t]{$j$}}
  \multiput(105,3)(0,40){2}{\makebox(5,5)[t]{$j$}}
  \multiput(10,63)(25,0){5}{\makebox(5,5)[t]{$k$}}
  \multiput(10,23)(25,0){5}{\makebox(5,5)[t]{$k$}}
  \multiput(110,3)(0,40){2}{\makebox(5,5)[t]{$k$}}
  \multiput(15,75)(25,0){4}{\vector(1,0){10}}
  \multiput(15,35)(25,0){4}{\vector(1,0){10}}
  \put(95,15){\line(0,1){20}}
  \put(95,55){\line(0,1){20}}
  \multiput(95,15)(0,40){2}{\vector(1,0){5}}
  \multiput(116.5,10.5)(0,20){4}{\makebox(5,5)[t]{=}}
  \multiput(114.5,15)(0,20){4}{\makebox(5,5)[t]{*}}
  \multiput(123,70)(5,0){4}{\line(0,1){10}}
  \multiput(123,70)(0,10){2}{\line(1,0){15}}
  \multiput(123,50)(5,0){4}{\line(0,1){10}}
  \multiput(123,50)(0,10){2}{\line(1,0){15}}
  \multiput(123,30)(5,0){4}{\line(0,1){10}}
  \multiput(123,30)(0,10){2}{\line(1,0){15}}
  \multiput(123,10)(5,0){4}{\line(0,1){10}}
  \multiput(123,10)(0,10){2}{\line(1,0){15}}
  \put(2.5,72.5){\vector(0,1){5}}
  \put(7.5,77.5){\vector(0,-1){5}}
  \put(12.5,77.5){\vector(0,-1){5}}
  \put(31.5,72.5){\vector(0,1){5}}
  \put(33.5,77.5){\vector(0,-1){5}}
  \put(37.5,77.5){\vector(0,-1){5}}
  \put(57.5,77.5){\vector(0,-1){5}}
  \put(61.5,72.5){\vector(0,1){5}}
  \put(63.5,77.5){\vector(0,-1){5}}
  \put(81.5,72.5){\vector(0,1){5}}
  \put(83.5,77.5){\vector(0,-1){5}}
  \put(87.5,77.5){\vector(0,-1){5}}
  \put(102.5,72.5){\vector(0,1){5}}
  \put(107.5,77.5){\vector(0,-1){5}}
  \put(112.5,77.5){\vector(0,-1){5}}
  \put(125.5,77.5){\vector(0,-1){5}}
  \put(130.5,72.5){\vector(0,1){5}}
  \put(136.5,77.5){\vector(0,-1){5}}
  \put(100,52.5){\makebox(5,5)[t]{$\Downarrow$}}
  \put(105,52.5){\makebox(5,5)[t]{$\Uparrow$}}
  \put(112.5,57.5){\vector(0,-1){5}}
  \put(123,52.5){\makebox(5,5)[t]{$\Uparrow$}}
  \put(128,52.5){\makebox(5,5)[t]{$\Downarrow$}}
  \put(135.5,57.5){\vector(0,-1){5}}
  \put(2.5,37.5){\vector(0,-1){5}}
  \put(7.5,32.5){\vector(0,1){5}}
  \put(12.5,37.5){\vector(0,-1){5}}
  \put(31.5,32.5){\vector(0,1){5}}
  \put(33.5,37.5){\vector(0,-1){5}}
  \put(37.5,37.5){\vector(0,-1){5}}
  \put(57.5,37.5){\vector(0,-1){5}}
  \put(61.5,32.5){\vector(0,1){5}}
  \put(63.5,37.5){\vector(0,-1){5}}
  \put(81.5,32.5){\vector(0,1){5}}
  \put(83.5,37.5){\vector(0,-1){5}}
  \put(87.5,37.5){\vector(0,-1){5}}
  \put(100,32.5){\makebox(5,5)[t]{$\Uparrow$}}
  \put(105,32.5){\makebox(5,5)[t]{$\Downarrow$}}
  \put(112.5,37.5){\vector(0,-1){5}}
  \put(123,32.5){\makebox(5,5)[t]{$\Downarrow$}}
  \put(130.5,37.5){\vector(0,-1){5}}
  \put(133,32.5){\makebox(5,5)[t]{$\Uparrow$}}
  \put(102.5,17.5){\vector(0,-1){5}}
  \put(107.5,12.5){\vector(0,1){5}}
  \put(112.5,17.5){\vector(0,-1){5}}
  \put(125.5,12.5){\vector(0,1){5}}
  \put(130.5,17.5){\vector(0,-1){5}}
  \put(135.5,17.5){\vector(0,-1){5}}
\end{picture}
\end{center}
\caption{
Schematic representation of the diagrams contributing to process [g] of Fig.~\ref{ord_4_3}.
}
\label{sg_4_3}
\end{figure}
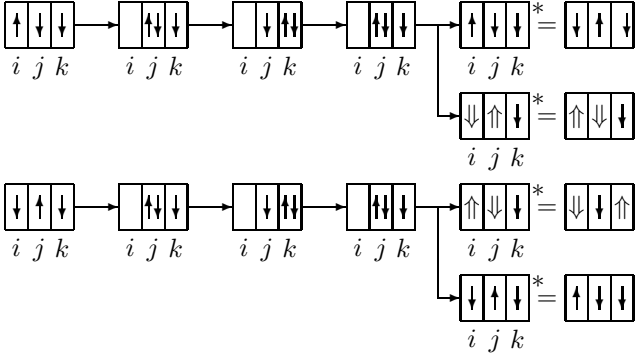

The processes [a]-[d] respectively [e] and [h] are equivalent to each other in each case since their individual steps are analogous. On account of the Pauli principle, contributions only result again for different spins on the sites $i$, $j$. Since time reversed initial states can be treated commonly by a spin summation, only two initial states must be examined. The diagrams contributing to the processes [a], [e], [f] and [g] are represented in the Fig. \ref{sa_4_3}, \ref{se_4_3}, \ref{sf_4_3} and \ref{sg_4_3}. The product of the pre-factors $D_{i,j}$ in accordance with Eq.~\eqref{G17} is invariant with respect to index exchanging of the form $j \leftrightarrow k$. Therefore such exchanging can also be practised in individual diagrams. The diagrams contributing to the process [g] of Fig.~\ref{ord_4_3} have a different pre-factor and are marked therefore in the Fig.~\ref{sg_4_3} by $*$. The pre-factor of these diagrams can be transformed into the form of the other diagrams by an index exchange $i \leftrightarrow j$.

Expression \eqref{G34} for the contribution ${\cal H}_{\rm eff}\big|_{\rm 4th\,ord.}^{3{\rm sites}}$ of the processes with three involved sites to the effective Hamiltonian can be written symbolically by using the discussed process classification and incidence
\begin{eqnarray}
  \lefteqn{{\cal H}_{\rm eff}\bigg|_{\rm 4th\,ord.}^{3\,{\rm sites}}U^{3}\;=}&&\nonumber\\
  &=&\phantom{\bigg|}+4\mbox{\setlength{\unitlength}{0.4ex}
\begin{picture}(15,12)
  {\small
  \multiput(0,-6)(15,0){2}{\line(0,1){15}}
  \multiput(0,-6)(0,7.5){3}{\line(1,0){15}}
  \put(0,-4.5){\makebox(5,5)[t]{$\uparrow$}}
  \put(5,-4.5){\makebox(5,5)[t]{$\downarrow$}}
  \put(10,-4.5){\makebox(5,5)[t]{$\downarrow$}}
  \put(0,3){\makebox(5,5)[t]{$\uparrow$}}
  \put(5,3){\makebox(5,5)[t]{$\downarrow$}}
  \put(10,3){\makebox(5,5)[t]{$\downarrow$}}
  }
\end{picture}}+4\mbox{\setlength{\unitlength}{0.4ex}
\begin{picture}(15,12)
  {\small
  \multiput(0,-6)(15,0){2}{\line(0,1){15}}
  \multiput(0,-6)(0,7.5){3}{\line(1,0){15}}
  \put(0,-4.5){\makebox(5,5)[t]{$\Downarrow$}}
  \put(5,-4.5){\makebox(5,5)[t]{$\downarrow$}}
  \put(10,-4.5){\makebox(5,5)[t]{$\Uparrow$}}
  \put(0,3){\makebox(5,5)[t]{$\uparrow$}}
  \put(5,3){\makebox(5,5)[t]{$\downarrow$}}
  \put(10,3){\makebox(5,5)[t]{$\downarrow$}}
  }
\end{picture}}+4\mbox{\setlength{\unitlength}{0.4ex}
\begin{picture}(15,12)
  {\small
  \multiput(0,-6)(15,0){2}{\line(0,1){15}}
  \multiput(0,-6)(0,7.5){3}{\line(1,0){15}}
  \put(0,-4.5){\makebox(5,5)[t]{$\Uparrow$}}
  \put(5,-4.5){\makebox(5,5)[t]{$\Downarrow$}}
  \put(10,-4.5){\makebox(5,5)[t]{$\downarrow$}}
  \put(0,3){\makebox(5,5)[t]{$\downarrow$}}
  \put(5,3){\makebox(5,5)[t]{$\uparrow$}}
  \put(10,3){\makebox(5,5)[t]{$\downarrow$}}
  }
\end{picture}}+4\mbox{\setlength{\unitlength}{0.4ex}
\begin{picture}(15,12)
  {\small
  \multiput(0,-6)(15,0){2}{\line(0,1){15}}
  \multiput(0,-6)(0,7.5){3}{\line(1,0){15}}
  \put(0,-4.5){\makebox(5,5)[t]{$\downarrow$}}
  \put(5,-4.5){\makebox(5,5)[t]{$\Downarrow$}}
  \put(10,-4.5){\makebox(5,5)[t]{$\Uparrow$}}
  \put(0,3){\makebox(5,5)[t]{$\downarrow$}}
  \put(5,3){\makebox(5,5)[t]{$\uparrow$}}
  \put(10,3){\makebox(5,5)[t]{$\downarrow$}}
  }
\end{picture}}+\nonumber\\
  &&\phantom{\bigg|}-2\mbox{\setlength{\unitlength}{0.4ex}
\begin{picture}(15,12)
  {\small
  \multiput(0,-6)(15,0){2}{\line(0,1){15}}
  \multiput(0,-6)(0,7.5){3}{\line(1,0){15}}
  \put(0,-4.5){\makebox(5,5)[t]{$\uparrow$}}
  \put(5,-4.5){\makebox(5,5)[t]{$\downarrow$}}
  \put(10,-4.5){\makebox(5,5)[t]{$\downarrow$}}
  \put(0,3){\makebox(5,5)[t]{$\uparrow$}}
  \put(5,3){\makebox(5,5)[t]{$\downarrow$}}
  \put(10,3){\makebox(5,5)[t]{$\downarrow$}}
  }
\end{picture}}-2\mbox{\setlength{\unitlength}{0.4ex}
\begin{picture}(15,12)
  {\small
  \multiput(0,-6)(15,0){2}{\line(0,1){15}}
  \multiput(0,-6)(0,7.5){3}{\line(1,0){15}}
  \put(0,-4.5){\makebox(5,5)[t]{$\Downarrow$}}
  \put(5,-4.5){\makebox(5,5)[t]{$\downarrow$}}
  \put(10,-4.5){\makebox(5,5)[t]{$\Uparrow$}}
  \put(0,3){\makebox(5,5)[t]{$\uparrow$}}
  \put(5,3){\makebox(5,5)[t]{$\downarrow$}}
  \put(10,3){\makebox(5,5)[t]{$\downarrow$}}
  }
\end{picture}}-2\mbox{\setlength{\unitlength}{0.4ex}
\begin{picture}(15,12)
  {\small
  \multiput(0,-6)(15,0){2}{\line(0,1){15}}
  \multiput(0,-6)(0,7.5){3}{\line(1,0){15}}
  \put(0,-4.5){\makebox(5,5)[t]{$\Uparrow$}}
  \put(5,-4.5){\makebox(5,5)[t]{$\Downarrow$}}
  \put(10,-4.5){\makebox(5,5)[t]{$\downarrow$}}
  \put(0,3){\makebox(5,5)[t]{$\downarrow$}}
  \put(5,3){\makebox(5,5)[t]{$\uparrow$}}
  \put(10,3){\makebox(5,5)[t]{$\downarrow$}}
  }
\end{picture}}-2\mbox{\setlength{\unitlength}{0.4ex}
\begin{picture}(15,12)
  {\small
  \multiput(0,-6)(15,0){2}{\line(0,1){15}}
  \multiput(0,-6)(0,7.5){3}{\line(1,0){15}}
  \put(0,-4.5){\makebox(5,5)[t]{$\downarrow$}}
  \put(5,-4.5){\makebox(5,5)[t]{$\Downarrow$}}
  \put(10,-4.5){\makebox(5,5)[t]{$\Uparrow$}}
  \put(0,3){\makebox(5,5)[t]{$\downarrow$}}
  \put(5,3){\makebox(5,5)[t]{$\uparrow$}}
  \put(10,3){\makebox(5,5)[t]{$\downarrow$}}
  }
\end{picture}}+\nonumber\\
  &&\phantom{\bigg|}-\phantom{2}\mbox{\setlength{\unitlength}{0.4ex}
\begin{picture}(15,12)
  {\small
  \multiput(0,-6)(15,0){2}{\line(0,1){15}}
  \multiput(0,-6)(0,7.5){3}{\line(1,0){15}}
  \put(0,-4.5){\makebox(5,5)[t]{$\uparrow$}}
  \put(5,-4.5){\makebox(5,5)[t]{$\downarrow$}}
  \put(10,-4.5){\makebox(5,5)[t]{$\downarrow$}}
  \put(0,3){\makebox(5,5)[t]{$\uparrow$}}
  \put(5,3){\makebox(5,5)[t]{$\downarrow$}}
  \put(10,3){\makebox(5,5)[t]{$\downarrow$}}
  }
\end{picture}}-\phantom{2}\mbox{\setlength{\unitlength}{0.4ex}
\begin{picture}(15,12)
  {\small
  \multiput(0,-6)(15,0){2}{\line(0,1){15}}
  \multiput(0,-6)(0,7.5){3}{\line(1,0){15}}
  \put(0,-4.5){\makebox(5,5)[t]{$\Downarrow$}}
  \put(5,-4.5){\makebox(5,5)[t]{$\downarrow$}}
  \put(10,-4.5){\makebox(5,5)[t]{$\Uparrow$}}
  \put(0,3){\makebox(5,5)[t]{$\uparrow$}}
  \put(5,3){\makebox(5,5)[t]{$\downarrow$}}
  \put(10,3){\makebox(5,5)[t]{$\downarrow$}}
  }
\end{picture}}-\phantom{2}\mbox{\setlength{\unitlength}{0.4ex}
\begin{picture}(15,12)
  {\small
  \multiput(0,-6)(15,0){2}{\line(0,1){15}}
  \multiput(0,-6)(0,7.5){3}{\line(1,0){15}}
  \put(0,-4.5){\makebox(5,5)[t]{$\Uparrow$}}
  \put(5,-4.5){\makebox(5,5)[t]{$\Downarrow$}}
  \put(10,-4.5){\makebox(5,5)[t]{$\downarrow$}}
  \put(0,3){\makebox(5,5)[t]{$\downarrow$}}
  \put(5,3){\makebox(5,5)[t]{$\uparrow$}}
  \put(10,3){\makebox(5,5)[t]{$\downarrow$}}
  }
\end{picture}}-\phantom{2}\mbox{\setlength{\unitlength}{0.4ex}
\begin{picture}(15,12)
  {\small
  \multiput(0,-6)(15,0){2}{\line(0,1){15}}
  \multiput(0,-6)(0,7.5){3}{\line(1,0){15}}
  \put(0,-4.5){\makebox(5,5)[t]{$\downarrow$}}
  \put(5,-4.5){\makebox(5,5)[t]{$\uparrow$}}
  \put(10,-4.5){\makebox(5,5)[t]{$\downarrow$}}
  \put(0,3){\makebox(5,5)[t]{$\downarrow$}}
  \put(5,3){\makebox(5,5)[t]{$\uparrow$}}
  \put(10,3){\makebox(5,5)[t]{$\downarrow$}}
  }
\end{picture}}+\nonumber\\
  &&\phantom{\bigg|}-\phantom{2}\mbox{\setlength{\unitlength}{0.4ex}
\begin{picture}(15,12)
  {\small
  \multiput(0,-6)(15,0){2}{\line(0,1){15}}
  \multiput(0,-6)(0,7.5){3}{\line(1,0){15}}
  \put(0,-4.5){\makebox(5,5)[t]{$\uparrow$}}
  \put(5,-4.5){\makebox(5,5)[t]{$\downarrow$}}
  \put(10,-4.5){\makebox(5,5)[t]{$\downarrow$}}
  \put(0,3){\makebox(5,5)[t]{$\uparrow$}}
  \put(5,3){\makebox(5,5)[t]{$\downarrow$}}
  \put(10,3){\makebox(5,5)[t]{$\downarrow$}}
  }
\end{picture}}-\phantom{2}\mbox{\setlength{\unitlength}{0.4ex}
\begin{picture}(15,12)
  {\small
  \multiput(0,-6)(15,0){2}{\line(0,1){15}}
  \multiput(0,-6)(0,7.5){3}{\line(1,0){15}}
  \put(0,-4.5){\makebox(5,5)[t]{$\Downarrow$}}
  \put(5,-4.5){\makebox(5,5)[t]{$\downarrow$}}
  \put(10,-4.5){\makebox(5,5)[t]{$\Uparrow$}}
  \put(0,3){\makebox(5,5)[t]{$\uparrow$}}
  \put(5,3){\makebox(5,5)[t]{$\downarrow$}}
  \put(10,3){\makebox(5,5)[t]{$\downarrow$}}
  }
\end{picture}}-\phantom{2}\mbox{\setlength{\unitlength}{0.4ex}
\begin{picture}(15,12)
  {\small
  \multiput(0,-6)(15,0){2}{\line(0,1){15}}
  \multiput(0,-6)(0,7.5){3}{\line(1,0){15}}
  \put(0,-4.5){\makebox(5,5)[t]{$\Uparrow$}}
  \put(5,-4.5){\makebox(5,5)[t]{$\Downarrow$}}
  \put(10,-4.5){\makebox(5,5)[t]{$\downarrow$}}
  \put(0,3){\makebox(5,5)[t]{$\downarrow$}}
  \put(5,3){\makebox(5,5)[t]{$\uparrow$}}
  \put(10,3){\makebox(5,5)[t]{$\downarrow$}}
  }
\end{picture}}-\phantom{2}\mbox{\setlength{\unitlength}{0.4ex}
\begin{picture}(15,12)
  {\small
  \multiput(0,-6)(15,0){2}{\line(0,1){15}}
  \multiput(0,-6)(0,7.5){3}{\line(1,0){15}}
  \put(0,-4.5){\makebox(5,5)[t]{$\downarrow$}}
  \put(5,-4.5){\makebox(5,5)[t]{$\uparrow$}}
  \put(10,-4.5){\makebox(5,5)[t]{$\downarrow$}}
  \put(0,3){\makebox(5,5)[t]{$\downarrow$}}
  \put(5,3){\makebox(5,5)[t]{$\uparrow$}}
  \put(10,3){\makebox(5,5)[t]{$\downarrow$}}
  }
\end{picture}}\label{G35}
\end{eqnarray}
The diagrams of Eq.~\eqref{G35} are to be interpreted again as hopping processes. Eq.~\eqref{G35} can be summarized formally
\begin{eqnarray}
   {\cal H}_{\rm eff}\bigg|_{\rm 4th\,ord.}^{3\,{\rm sites}}&=&\frac{2}{U^{3}}\left[\mbox{\setlength{\unitlength}{0.4ex}
\begin{picture}(15,12)
  {\small
  \multiput(0,-6)(15,0){2}{\line(0,1){15}}
  \multiput(0,-6)(0,7.5){3}{\line(1,0){15}}
  \put(0,-4.5){\makebox(5,5)[t]{$\downarrow$}}
  \put(5,-4.5){\makebox(5,5)[t]{$\Downarrow$}}
  \put(10,-4.5){\makebox(5,5)[t]{$\Uparrow$}}
  \put(0,3){\makebox(5,5)[t]{$\downarrow$}}
  \put(5,3){\makebox(5,5)[t]{$\uparrow$}}
  \put(10,3){\makebox(5,5)[t]{$\downarrow$}}
  }
\end{picture}}-\mbox{\setlength{\unitlength}{0.4ex}
\begin{picture}(15,12)
  {\small
  \multiput(0,-6)(15,0){2}{\line(0,1){15}}
  \multiput(0,-6)(0,7.5){3}{\line(1,0){15}}
  \put(0,-4.5){\makebox(5,5)[t]{$\downarrow$}}
  \put(5,-4.5){\makebox(5,5)[t]{$\uparrow$}}
  \put(10,-4.5){\makebox(5,5)[t]{$\downarrow$}}
  \put(0,3){\makebox(5,5)[t]{$\downarrow$}}
  \put(5,3){\makebox(5,5)[t]{$\uparrow$}}
  \put(10,3){\makebox(5,5)[t]{$\downarrow$}}
  }
\end{picture}}\right]\label{G36}
\end{eqnarray}
and transferred to an operator equation 
\begin{eqnarray}
  {\cal H}_{\rm eff}\bigg|_{\rm 4th\,ord.}^{3\,{\rm sites}}&=&-2\frac{t^{4}}{U^{3}}\sum\limits_{i,j,k,\sigma}D_{i,j}D_{j,i}D_{i,k}D_{k,i}\cdot\nonumber\\
  &&\cdot\left[ n_{i,\sigma}c^{\dagger}_{j,\sigma}c_{j,-\sigma}c_{k,\sigma}c^{\dagger}_{k,-\sigma}+n_{i,\sigma}n_{j,-\sigma}n_{k,\sigma}\right].\nonumber\\
  &&\label{G37}
\end{eqnarray}
Thereby the lattice sites summations, the spin summation and the pre-factors have to be considered. The sign of the diagrams results again from the special process execution and the fermionic anti-commutation relations. Eq.~\eqref{G37} can be summarized
\begin{eqnarray}
  {\cal H}_{\rm eff}\bigg|_{\rm 4th\,ord.}^{3\,{\rm sites}}&=&2\frac{t^{4}}{U^{3}}\sum\limits_{i,j,k}D_{i,j}D_{j,i}D_{i,k}D_{k,i}\left[{\bf S}_{j}\cdot{\bf S}_{k}-\frac{1}{4}\right]\nonumber\\
  &&\label{G38}
\end{eqnarray}
using spin operators \eqref{G31}.

If the processes with two and four involved sites are also treated in this form, all contributions to the fourth-order perturbation are considered and up to fourth order one obtains the effective Hamiltonian
%
%
%
%
\end{multicols}
\widetext
\begin{eqnarray}
  {\cal H}_{\rm eff}&=&2\frac{t^{2}}{U}\sum\limits_{i,j}D_{i,j}D_{j,i}\left[{\bf S}_{i}\cdot{\bf S}_{j}-\frac{1}{4}\right]+\nonumber\\
  &&+\frac{t^{4}}{U^{3}}\Bigg[-8\sum\limits_{i,j}D_{i,j}^{2}D_{j,i}^{2}\left\{{\bf S}_{i}\cdot{\bf S}_{j}-\frac{1}{4}\right\}+2\sum\limits_{i,j,k}D_{i,j}D_{j,i}D_{i,k}D_{k,i}\left\{{\bf S}_{j}\cdot{\bf S}_{k}-\frac{1}{4}\right\}+\nonumber\\
  &&\qquad+\frac{1}{2}\sum\limits_{i,j,k,l}D_{i,j}D_{j,k}D_{k,l}D_{l,i}\left\{ \frac{1}{4}-{\bf S}_{i}\cdot{\bf S}_{j}-{\bf S}_{i}\cdot{\bf S}_{k}-{\bf S}_{i}\cdot{\bf S}_{l}-{\bf S}_{j}\cdot{\bf S}_{k}-{\bf S}_{j}\cdot{\bf S}_{l}-{\bf S}_{k}\cdot{\bf S}_{l}+\right.\nonumber\\
  &&\qquad\qquad +20\left({\bf S}_{i}\cdot{\bf S}_{j}\right)\left({\bf S}_{k}\cdot{\bf S}_{l}\right)+20\left({\bf S}_{i}\cdot{\bf S}_{l}\right)\left({\bf S}_{j}\cdot{\bf S}_{k}\right)-20\left({\bf S}_{i}\cdot{\bf S}_{k}\right)\left({\bf S}_{j}\cdot{\bf S}_{l}\right)\bigg\}\Bigg].\label{G39}
\end{eqnarray}
%
%
%
%
%
%
%
%
%
\begin{multicols}{2}
\narrowtext
This result is valid for arbitrary lattices and dimensions since all such information is contained in the hopping matrix $D_{i,j}$. The derived fourth-order perturbation of the effective Hamiltonian agrees with other results which have been obtained using canonical transformations \cite{MacDonald,Stein} or projection technique \cite{Takahashi1}. 

In the case of the linear chain with nearest-neighbor hopping, processes with four involved sites do not contribute to the fourth-order perturbation. Therefore in this case Eq.~\eqref{G39} coincides with the result of Klein and Seitz \cite{Klein1}.


\section{Conclusions}\label{Conclusions}

In this paper we have presented a cumulant approach for the construction of effective Hamiltonians. The size consistency of the results is always guaranteed by the introduction of generalized cumulant expressions. While a previous cumulant method \cite{Polatsek1} has been limited to systems containing two interacting subsystems, the formalism presented here can be applied to systems with arbitrary Hilbert space for any temperature.

We have applied the presented cumulant method to the half-filled Hubbard model on a general lattice in arbitrary spatial dimension for which the fourth-order perturbation expansion of the effective Hamiltonian was calculated. In the past the fourth-order perturbation was controversially discussed and, therefore, it can be considered as a reference problem of methods for the construction of effective Hamiltonians. The discussion of the strong-coupling expansion for the Hubbard model demonstrates the power of the derived cumulant approach, it enables a transparent construction of size-consistent effective Hamiltonians for systems with arbitrary Hilbert space. The derived fourth-order perturbation of the Hubbard model agrees with results of other methods.


\acknowledgments
{
Support from the Deutsche Forschungsgemeinschaft is gratefully acknowledged. The authors thank C. Weidacher for helpful discussions.
}


\end{multicols}
\widetext

\begin{appendix}

\section{Generalized cumulants}\label{A1}

The cumulant expressions defined by Eq.~\eqref{G8a} are to be interpreted as a generalization of cumulant expectation values \cite{Kubo,Kladko}. In the Eq.~\eqref{G8a}, the $X_{i}$ are operators whose powers of $\nu_{i}$ are to be considered in the cumulant expansion. $P$ denotes a projector onto a subspace of the system. Generalized cumulants are traced back to the definition \eqref{G8a}, i.e. forming cumulant expressions of operator functions are to be explained by means of power series. For computing generalized cumulants, a function of non-commuting operators is to be differentiated in accordance with Eq.~\eqref{G8a}. Therefore, these cumulant expressions must be calculated with the help of series expansions.

From Eq.~\eqref{G8a} follows by expanding the exponential function
%
%
%
%
\begin{eqnarray}
  \left(\prod_{i=1}^{N} X^{\nu_{i}}_{i}\right)^{C}_{P}&=&\left.\left[\prod_{i=1}^{N}\left(\frac{\partial}{\partial \xi_{i}}\right)^{\nu_{i}}\right]\left[\ln P+\ln\left\{ 1+\sum_{\genfrac{}{}{0pt}{1}{n_{1},\dots,n_{N}=0}{(n_{1},\dots,n_{N})\not=(0,\dots,0)}}^{\infty}\frac{\xi_{1}^{n_{1}}}{n_{1}!}\dots\frac{\xi_{N}^{n_{N}}}{n_{N}!}\left( X^{n_{1}}_{1}\dots X^{n_{N}}_{N}\right)_{P}\right\}\right]\right|_{\xi_{i}=0\;\forall i}\label{G40}
\end{eqnarray}
%
%
%
%
%
%
where the properties of projection operators were utilized. The first addend of Eq.~\eqref{G40} does not contribute since it is still to be differentiated. The residual logarithm has the form $\ln(1+x)$ and therefore, can be expanded as follows:
\begin{eqnarray}
  \left(\prod_{i=1}^{N} X^{\nu_{i}}_{i}\right)^{C}_{P}&=&\left[\prod_{i=1}^{N}\left(\frac{\partial}{\partial \xi_{i}}\right)^{\nu_{i}}\right]\left[\left\{\sum_{\genfrac{}{}{0pt}{1}{n_{1},\dots,n_{N}=0}{(n_{1},\dots,n_{N})\not=(0,\dots,0)}}^{\infty}\mbox{\hspace{-0.7cm}}\frac{\xi_{1}^{n_{1}}}{n_{1}!}\dots\frac{\xi_{N}^{n_{N}}}{n_{N}!}\left( X^{n_{1}}_{1}\dots X^{n_{N}}_{N}\right)_{P}\right\}+\right.\nonumber\\
  &&-\frac{1}{2}\left\{\sum_{\genfrac{}{}{0pt}{1}{\genfrac{}{}{0pt}{1}{n_{1},\dots,n_{N}=0}{m_{1},\dots,m_{N}=0}}{\genfrac{}{}{0pt}{1}{(n_{1},\dots,n_{N})\not=(0,\dots,0)}{(m_{1},\dots,m_{N})\not=(0,\dots,0)}}}^{\infty}\mbox{\hspace{-0.7cm}}\frac{\xi_{1}^{(n_{1}+m_{1})}}{n_{1}!m_{1}!}\dots\frac{\xi_{N}^{(n_{N}+m_{N})}}{n_{N}!m_{N}!}\left( X_{1}^{n_{1}}\dots X_{N}^{n_{N}}\right)_{P}\left( X_{1}^{m_{1}}\dots X_{N}^{m_{N}}\right)_{P}\right\}\nonumber\\
  &&+\frac{1}{3}\Bigg\{\dots\Bigg\}+\dots\Bigg]\Bigg|_{\xi_{i}=0\;\forall i}.\label{G41}
\end{eqnarray}
If one differentiates the sums of Eq.~\eqref{G41} one obtains:
%
%
%
%
\begin{eqnarray}
  \left( X^{\nu_{1}}_{1}\dots X^{\nu_{N}}_{N}\right)^{C}_{P}&=&\left( X_{1}^{\nu_{1}}\dots X_{N}^{\nu_{N}}\right)_{P}-\frac{1}{2}\sum_{\genfrac{}{}{0pt}{1}{\genfrac{}{}{0pt}{1}{n_{1},\dots,n_{N}=0}{m_{1},\dots,m_{N}=0}}{\genfrac{}{}{0pt}{1}{(n_{1},\dots,n_{N})\not=(0,\dots,0)}{(m_{1},\dots,m_{N})\not=(0,\dots,0)}}}^{\infty}\mbox{\hspace{-0.7cm}}\delta(\nu_{1},n_{1}+m_{1})\dots\delta(\nu_{N},n_{N}+m_{N})\frac{\nu_{1}!}{n_{1}!m_{1}!}\dots\frac{\nu_{N}!}{n_{N}!m_{N}!}\cdot\nonumber\\
  &&\mbox{\hspace{7cm}}\cdot\left( X_{1}^{n_{1}}\dots X_{N}^{n_{N}}\right)_{P}\left( X_{1}^{m_{1}}\dots X_{N}^{m_{N}}\right)_{P}\qquad+\dots\label{G42}
\end{eqnarray}
%
%
%
%
%
%
From Eq.~\eqref{G42} one can obtain special cumulant expressions. For example, one finds:
\begin{eqnarray}
  (A)_{P}^{C}&=&(A)_{P},\label{G43}\\
  \left(A^{2}\right)_{P}^{C}&=&\left(A^{2}\right)_{P}-(A)_{P}^{2},\label{G44}\\
  (AB)^{C}_{P}&=&(AB)_{P}-\frac{1}{2}(A)_{P}(B)_{P}-\frac{1}{2}(B)_{P}(A)_{P}.\label{G45}
\end{eqnarray}

\end{appendix}


\begin{multicols}{2}
\narrowtext

\end{multicols}
\widetext














\end{document}